\documentclass[journal]{IEEEtran}
\usepackage{amsmath}
\usepackage{amssymb}
\usepackage{bm}
\usepackage{graphicx}
\usepackage{booktabs}
\usepackage{url}
\usepackage{cite}

\begin{document}
\bstctlcite{IEEEexample:BSTcontrol} 

\title{Broadband Characterization of Polarization Mode Dispersion for Quantum Communication Channels}

\author{Vadim~Rodimin,
        Konstantin~Kravtsov,
        Rui~Ming~Chua,
        Xingjian~Zhang,
        Aleksei~Ponasenko,
        Yury~Kurochkin,
        Alexander~Ling,
        and~James~A.~Grieve%
\thanks{This work was supported by Abu Dhabi's Advanced Technology Research Council. (Corresponding author: Vadim Rodimin.)}%
\thanks{V. Rodimin, K. Kravtsov, X. Zhang, A. Ponasenko, Y. Kurochkin, and J. A. Grieve are with the Quantum Research Center, Technology Innovation Institute, Masdar City, Abu Dhabi, UAE (e-mail: vadim.rodimin@tii.ae).}%
\thanks{R. M. Chua is with the Quantum Research Center, Technology Innovation Institute, Masdar City, Abu Dhabi, UAE, and also with the Centre for Quantum Technologies, National University of Singapore, 3 Science Drive 2, 117543 Singapore.}%
\thanks{A. Ling is with the Centre for Quantum Technologies, National University of Singapore, 3 Science Drive 2, 117543 Singapore, and also with the Department of Physics, National University of Singapore, Blk S12, 2 Science Drive 3, 117551 Singapore.}}

\maketitle

\begin{abstract}
We present a method for characterizing depolarization in fiber channels carrying broadband polarization-encoded optical signals. Wavelength-dependent polarization mode dispersion maps each input polarization to a trajectory on the Poincar\'e sphere; we show that the singular value decomposition of the band-averaged rotation matrix yields, in closed form, the optimal input states, the associated mutually unbiased measurement bases, and the resulting projection error probabilities. The three singular values provide a compact, bandwidth-dependent channel signature that distinguishes first- from higher-order dispersion, and yield the bandwidth at which the worst-case error probability reaches 3\% as a practical filtering budget. We perform a second-order analysis of the band average that treats the two higher-order mechanisms: variation of the differential group delay at a fixed principal-state axis and wander of that axis with wavelength. We characterize 12 deployed fiber links in Masdar City and demonstrate mitigation by concatenating two channels through a single polarization controller, a step towards nonlocal compensation. The analysis is implemented in an openly available Python library for fiber-channel characterization.
\end{abstract}


\section{Introduction}
Quantum networks require faithful transmission of polarization-encoded quantum states, driving the development of diverse photonic sources. Unlike the narrowband sources of conventional telecommunication, many quantum sources are intrinsically broadband~\cite{yadav,Almassri:25,zhou2018room}, while quantum protocols demand preservation of the qubit state over the full spectrum. This makes wavelength-dependent polarization distortions a critical limitation for quantum-state transmission.

Narrowband filtering reduces polarization distortion but sacrifices photon flux, which cannot always be recovered by increasing source brightness~\cite{du2025entanglement,du2024demonstration}. The discarded bandwidth may also carry useful quantum information, for example, in polarization--time-energy hyperentanglement~\cite{Chen:17}, making broadband preservation desirable. Preserving the full bandwidth therefore requires understanding how the fiber transforms polarization across wavelength.

Given a fixed polarization input, polarization mode dispersion (PMD) causes an output polarization state $\bm{s}_{\mathrm{out}}$ to evolve with the wavelength, forming a particular trajectory on the Poincar\'e sphere~\cite{galtarossa,gordon-kogelnik}. Classical coherent receivers recover the full optical field and can compensate wavelength-dependent polarization distortions digitally. Quantum receivers cannot perform such post-detection inversion: measurement is destructive, and protocols require two mutually unbiased bases to be preserved at once. In many practical applications, such as quantum key distribution (QKD), the whole trajectory is projected onto a single fixed measurement vector, and the resulting error rate is quantified by the {\em infidelity}~\cite{ballance}. Conventional telecom PMD metrics based on a single differential group delay (DGD) value~\cite{itu-g652}, with fixed principal states of polarization (PSP), are insufficient to characterize broadband polarization evolution or to select optimal bases. While optimal basis selection has been addressed in quantum settings~\cite{rodimin-pra,shtaif-oe,nsr2025}, the treatment remains within a first-order approximation.

Following the analysis of PMD-induced decoherence of two-photon entanglement and its nonlocal compensation~\cite{antonelli-prl,shtaif-oe}, the present work addresses a complementary question: given the measured wavelength-resolved transformation of one deployed fiber, which measurement bases minimize the broadband projection errors? We develop a quantum-specific metrological framework that describes general polarization evolution with higher-order PMD, which varies across deployed fibers and becomes increasingly important over broad bandwidths. It is used to analyze the impact of higher-order PMD, derive practical metrics of channel characterization in the vicinity of a reference wavelength, and balance the count rate against the level of depolarization for a given channel. A broadband quantum channel is therefore characterized not by one number but by its action on a complete set of bases.

\section{Analysis of PMD-Related Measurement Errors}
\label{sec:framework}
In this section, we analyze the polarization states entering and exiting a fiber channel. All polarizations are represented as Stokes vectors and can be visualized using the Poincar\'e sphere.

In the case of first-order PMD, the evolution of the output polarization with changing wavelength is equivalent to a rotation around the PSP. The PSP axis thus defines the poles and equator of this rotation. However, real channels require a more comprehensive description.

\subsection{Probability of Projection Errors}

The channel polarization transformation $\mathsf{R}(\lambda)$ as a function of wavelength $\lambda$ inside the chosen wavelength band is defined by
\begin{equation}
\bm{s}_{\mathrm{out}}(\bm{s}_{\mathrm{in}},\lambda)
= \mathsf{R}(\lambda)\,\bm{s}_{\mathrm{in}},
\label{eq:transformation}
\end{equation}
where $\bm{s}$ is a unit Stokes vector and $\mathsf{R}(\lambda)\in SO(3)$ is a wavelength-dependent rotation of the Poincar\'e sphere. The rotation is universal for the whole sphere: each input state traces its own trajectory, but all trajectories are orbits of the same $\mathsf{R}(\lambda)$. Equation~(\ref{eq:transformation}) assumes a unitary (lossless, depolarization-free) channel at every $\lambda$, so that $\mathsf{R}(\lambda)$ is a proper rotation.

We quantify the probability of projection errors of a trajectory, $p_e$, by the infidelity of the band-filtered light with respect to a fixed measurement vector $\bm{e}$: the integral below is the fidelity of the filtered state with the measurement state, and $p_e$ is its complement,
\begin{equation} \label{eq:arc_int}
p_e(\bm{e})
= 1 - \frac{1}{\Delta\lambda}
\int_{\Delta\lambda}
\bigl|\langle s_{\mathrm{out}}(\lambda)|e\rangle\bigr|^2\, d\lambda,
\end{equation}
where the squared norm of a scalar product can be expressed in terms of the Stokes vectors $\bm{s}_{\mathrm{out}}$ and $\bm{e}$ as $|\langle s_{\mathrm{out}}|e\rangle|^2 = \tfrac{1}{2}(1+\bm{s}_{\mathrm{out}}\cdot\bm{e})$. Carrying out the averaging over $\lambda$, we notice that (\ref{eq:arc_int}) is \emph{linear} in $\bm{e}$,
\begin{equation}
p_e(\bm{e})
= \frac{1}{2} - \frac{1}{2}\,\bm{e}\cdot\langle\bm{s}_{\mathrm{out}}\rangle_\lambda,
\quad
\langle\bm{s}_{\mathrm{out}}\rangle_\lambda
= \frac{1}{\Delta\lambda}\int \bm{s}_{\mathrm{out}}(\lambda)\,d\lambda ,
\label{eq:pe_linear}
\end{equation}
so that for any trajectory the optimal unit projector -- the minimizer of~(\ref{eq:pe_linear}) -- is the normalized chordal mean $\bm{e}=\langle\bm{s}_{\mathrm{out}}\rangle_\lambda/|\langle\bm{s}_{\mathrm{out}}\rangle_\lambda|$, and the corresponding minimal $p_e$ is
\begin{equation}
p_e = \frac{1 - |\langle\bm{s}_{\mathrm{out}}\rangle_\lambda|}{2}.
\label{eq:pe_exact}
\end{equation}
The vector $\langle\bm{s}_{\mathrm{out}}\rangle_\lambda$ is an average of distinct unit vectors and therefore lies inside the sphere; its length equals the degree of polarization (DOP) of the band-filtered light, so (\ref{eq:pe_exact}) reads $p_e = (1-\mathrm{DOP})/2$. Because the Bloch vector of the band-filtered state is parallel to the matched projector, $p_e$ is at the same time the trace distance between the filtered state and the measurement state~\cite{nielsen-chuang}; the two standard distance measures coincide for the optimal projector, though not for a misaligned one.

\subsection{The Band-Averaged Rotation Matrix}
\label{subsec:mmatrix}

Our central object is closely related to the averaged Mueller matrix of classical depolarization optics, whose polar decomposition into retarder and depolarizer factors is a standard data-reduction tool~\cite{lu-chipman}. Define the band-averaged rotation matrix
\begin{equation}
\mathsf{M}
= \langle\mathsf{R}(\lambda)\rangle_\lambda
= \frac{1}{\Delta\lambda}\int \mathsf{R}(\lambda)\,d\lambda .
\label{eq:Mdef}
\end{equation}
$\mathsf{M}$ is a single $3\times3$ matrix. Being an average of rotations rather than a rotation~\cite{moakher}, it is a contraction: its singular values fall below unity in proportion to how much $\mathsf{R}(\lambda)$ varies across the band. A flat spectral density across the band is assumed throughout; a non-flat source spectrum $S(\omega)$ is accommodated by the weighted average $\mathsf{M}=\int S\,\mathsf{R}\,d\omega/\int S\,d\omega$, with all results below unchanged in form. The chordal mean of the trajectory launched from $\bm{s}_{\mathrm{in}}$ is, by linearity,
\begin{equation}
\langle\bm{s}_{\mathrm{out}}\rangle_\lambda
= \langle\mathsf{R}(\lambda)\bm{s}_{\mathrm{in}}\rangle_\lambda
= \mathsf{M}\,\bm{s}_{\mathrm{in}},
\end{equation}
so (\ref{eq:pe_exact}) gives the optimized $p_e$ of that trajectory directly in terms of the input,
\begin{equation}
p_e(\bm{s}_{\mathrm{in}})
= \frac{1 - |\mathsf{M}\,\bm{s}_{\mathrm{in}}|}{2}.
\label{eq:pe_M}
\end{equation}
Finding the optimal input states, together with their measurement bases, reduces entirely to the singular value decomposition (SVD) of the averaged rotation matrix~\cite{ossikovski-svd,heffner}
\begin{equation}
\mathsf{M} = \sum_{i=1}^{3}\sigma_i\,\bm{u}_i\bm{v}_i^{\!\top},
\qquad
\sigma_1 \ge \sigma_2 \ge \sigma_3 \ge 0,
\label{eq:svd}
\end{equation}
with orthonormal right- and left-singular vectors $\{\bm{v}_i\}$, $\{\bm{u}_i\}$. The three singular values, which we call a $\sigma$-signature of the channel, coincide with the canonical parameters of the unital qubit channel induced by the band averaging~\cite{ruskai}; they provide a basis-independent, three-number characterization of the channel, here measured as a function of the filtering bandwidth. Geometrically, $\mathsf{M}$ maps the Poincar\'e sphere onto an ellipsoid with semi-axes $\sigma_i\bm{u}_i$, the Stokes-space counterpart of the Bloch-ellipsoid picture of a qubit channel~\cite{nielsen-chuang}. Note that the decomposition acts on the averaged map, not on the point cloud of a single measured trajectory; a principal-component analysis of the latter optimizes a second-moment functional and differs beyond first order.

\subsection{Polar Trajectory}
\label{subsec:polar}

By (\ref{eq:pe_M}), the smallest $p_e$ is obtained for the input that $\mathsf{M}$ contracts the least, i.e., the leading right-singular vector of the decomposition~(\ref{eq:svd}). We define the polar trajectory $\Gamma_p$ as the most compact trajectory on the sphere generated by the input state $\bm{s}_p$
\begin{equation}
\bm{s}_p = \bm{v}_1,
\qquad
p_e^{\mathrm{pol}} = \frac{1-\sigma_1}{2},
\label{eq:polar_svd}
\end{equation}
which is the precise meaning of ``most compact'': the minimum $p_e$ (maximum-DOP) trajectory over the interval $\Delta\lambda$ (a DOP-based counterpart of the delay-based generalized principal states of optical pulses~\cite{shieh}). A minimizer always exists; it is unique up to a sign when $\sigma_1 > \sigma_2$. In the case of degenerate singular values, the corresponding optimal input polarization may be arbitrarily chosen within the degenerate subspace.

\begin{figure}[!t]
\centering
\includegraphics[width=0.8\columnwidth]{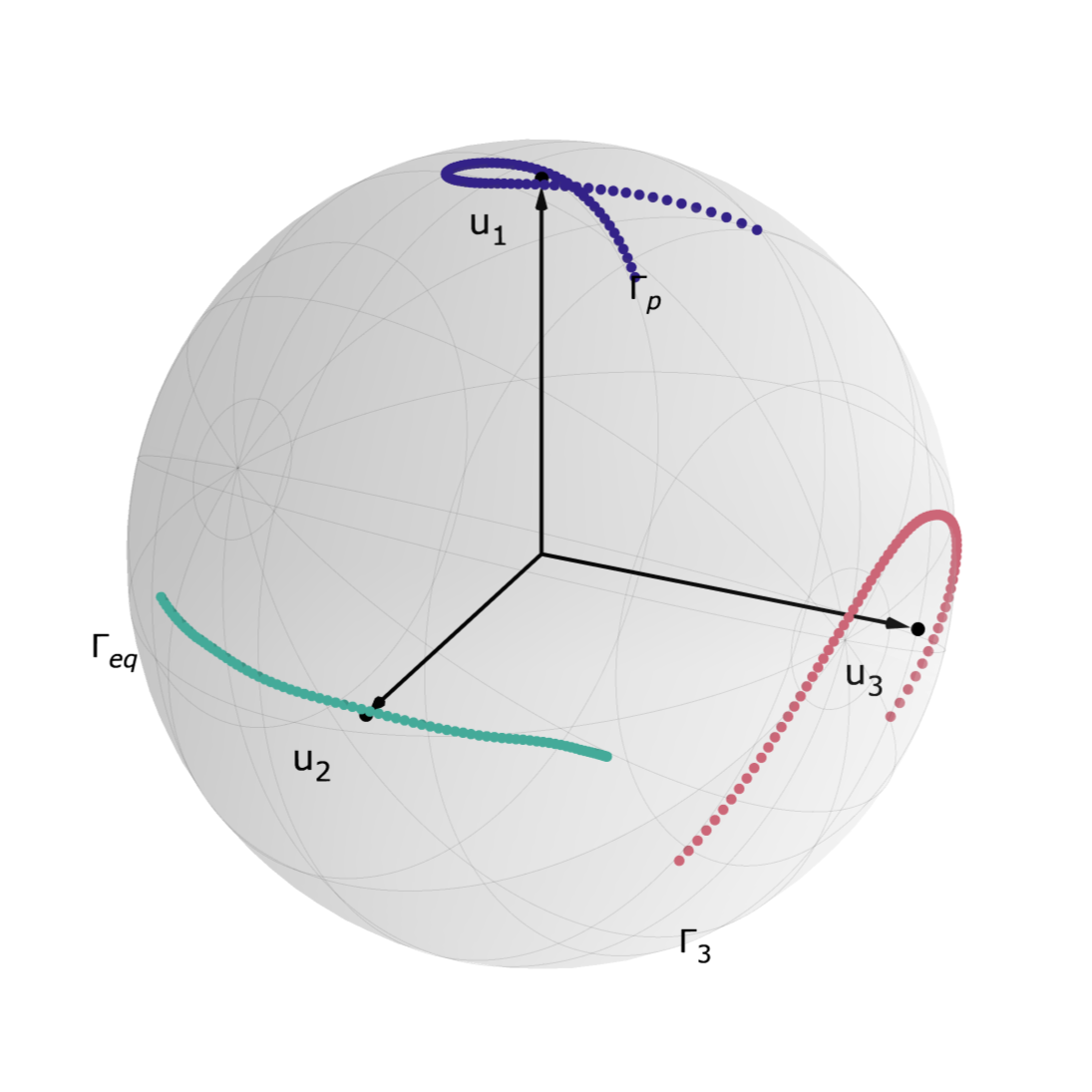}
\caption{Wavelength-dependent polarization transformation maps every input state to its own trajectory on the Poincar\'e sphere. The channel singles out three of them: the most compact polar one $\Gamma_p$, the equatorial $\Gamma_{\mathrm{eq}}$, and the most spread, worst-case $\Gamma_3$. Their launch states and the matched projectors $\bm{u}_1$, $\bm{u}_2$, $\bm{u}_3$ --- mutually perpendicular on the sphere, hence mutually unbiased as measurement bases --- follow from the singular value decomposition~(\ref{eq:svd}). Data processed from polarization measurements of a 10~km deployed fiber channel (Channel~4), Masdar City, Abu Dhabi, over 1300--1320~nm; the singular values are $0.974$, $0.909$, and $0.885$, respectively.}
\label{fig:trajectories}
\end{figure}

The optimal measurement vector for $\Gamma_p$ follows from the same decomposition. Since $\mathsf{M}\bm{v}_1 = \sigma_1\bm{u}_1$, the chordal mean of the polar trajectory points along $\bm{u}_1$, so the optimal projector is
\begin{equation}
\bm{e}_p = \frac{\mathsf{M}\bm{s}_p}{|\mathsf{M}\bm{s}_p|} = \bm{u}_1 ,
\label{eq:proj_svd}
\end{equation}
see Fig.~\ref{fig:trajectories}. The measurement basis for this trajectory is $\{\bm{e}_p,-\bm{e}_p\}$, $-\bm{e}_p$ being the orthogonal outcome (antipodal on the sphere) whose trajectory is the central reflection of $\Gamma_p$.

\subsection{Equatorial Trajectories}

For mutually unbiased quantum measurements, as in the commonly used BB84 QKD protocol, we need a second measurement basis whose Stokes projector is perpendicular to $\bm{e}_p$ on the Poincar\'e sphere; perpendicular Stokes vectors correspond to mutually unbiased -- not orthogonal -- quantum states. The singular value decomposition supplies such a basis automatically: both singular bases are orthonormal, so taking the second input state to be $\bm{s}_{\mathrm{eq}} = \bm{v}_2$, perpendicular to $\bm{s}_p$ on the sphere, gives the equatorial trajectory $\Gamma_{\mathrm{eq}}$ with optimal projector
\begin{equation}
\bm{e}_{\mathrm{eq}} = \bm{u}_2 ,
\qquad
\bm{e}_{\mathrm{eq}} \perp \bm{e}_p .
\end{equation}
The remaining singular direction $\bm{v}_3$ (projector $\bm{u}_3$) generates the trajectory $\Gamma_3$, which carries the largest error probability, $(1-\sigma_3)/2$, and provides the third mutually unbiased basis at no extra cost.

Limiting cases make this construction concrete. For pure first-order PMD the transverse plane is isotropic about the PSP, so $\sigma_2=\sigma_3$; the polar state is the PSP itself ($\sigma_1=1$, zero $p_e$), and the equatorial choice is immaterial. Higher-order PMD may lift this degeneracy, $\sigma_2>\sigma_3$, and selects a definite $\bm{s}_{\mathrm{eq}}$. In the case of isotropic depolarization all singular values are equal, so the choice of the basis does not affect the observed projection errors.

\subsection{Symmetric Trajectory}
\label{subsec:symmetric}

The polar and equatorial trajectories realize the smallest and the second-smallest error probabilities, $(1-\sigma_1)/2$ and $(1-\sigma_2)/2$; the largest one belongs to the third singular direction, $p_e=(1-\sigma_3)/2$. It is natural to ask for the best pair of mutually unbiased bases (MUB): the orthonormal input pair $(\bm{a},\bm{b})$ that minimizes the average error probability $\tfrac12[p_e(\bm{a})+p_e(\bm{b})]$, equivalently maximizes $|\mathsf{M}\bm{a}|+|\mathsf{M}\bm{b}|$.

For any orthonormal vectors $\bm{a}$, $\bm{b}$ and $\bm{c}$, the total squared stretch is invariant (invariance of the Frobenius norm): $|\mathsf{M}\bm{a}|^2+|\mathsf{M}\bm{b}|^2+|\mathsf{M}\bm{c}|^2 = \sigma_1^2+\sigma_2^2+\sigma_3^2$. Maximizing the sum over the pair is therefore equivalent to minimizing $|\mathsf{M}\bm{c}|^2$, which is the smallest for $\bm{c}=\bm{v}_3$, so the optimal pair $(\bm{a},\bm{b})$ lies in the $\bm{s}_p$--$\bm{s}_{\mathrm{eq}}$ plane. Since the square root is concave, the sum of the two moduli is maximal when they are equal. The balanced pair is $(\bm{s}_p\pm\bm{s}_{\mathrm{eq}})/\sqrt{2}$ -- the two input states rotated by $45^\circ$; their two trajectories share one and the same $p_e$,
\begin{equation}
p_e(\bm{a}) = p_e(\bm{b}) = \bar{p}_e^{\,\mathrm{opt}}
= \frac{1 - \sqrt{(\sigma_1^2+\sigma_2^2)/2}}{2}.
\label{eq:mub_opt}
\end{equation}
Minimizing the average thus automatically equalizes the two bases; the optimal pair is symmetric, which gives this subsection its name. The pole/equator average $\tfrac14(2-\sigma_1-\sigma_2)$ exceeds the optimum~(\ref{eq:mub_opt}) only at fourth order in the trajectory spread; in our measured channels the $45^\circ$ average coincides with the pole/equator average (see, e.g., Fig.~\ref{fig:combined}). We therefore report the latter as a convenient proxy for the optimal MUB-pair $p_e$, (\ref{eq:mub_opt}).

\subsection{Filtering Budget}

The framework naturally provides a scalar descriptor of the usable spectral bandwidth. We define a \emph{filtering budget} as the bandwidth over which the worst-case projection error remains below a prescribed threshold. The term is used in the same sense as the power and loss budgets of classical link design and the error budgets of quantum gate experiments~\cite{ballance}: bandwidth is spent against a fixed error limit.

Consider the asymptotic BB84 secret-key rate with key fraction $1-2h(Q)$, where $h$ is the binary Shannon entropy and $Q$ the quantum bit error rate (QBER)~\cite{shor-preskill}. We write
\begin{equation}
Q(\Delta\lambda)=a\,(\Delta\lambda)^2+q_0,
\label{eq:qber_model}
\end{equation}
where the quadratic term describes the PMD-related contribution to the QBER~\cite{rodimin-pra}, while $q_0$ collects bandwidth-independent errors. Since the photon flux scales as $\Delta\lambda$, the corresponding key rate is
\begin{equation}
K_r \propto \Delta\lambda\bigl[1-2h\bigl(Q(\Delta\lambda)\bigr)\bigr].
\label{eq:key_rate}
\end{equation}
Maximizing $K_r$ over $\Delta\lambda$ places the optimum where the PMD-related error contribution $a(\Delta\lambda)^2$ is close to $3\%$: it shifts only from $3.0\%$ at $q_0=0$ to $2.6\%$ at $q_0=2\%$, and is independent of $a$. The $3\%$ level thus reflects the rate--error trade-off rather than the properties of a particular channel or receiver. This motivates defining the channel filtering budget $\Delta\lambda_{3\%}$ as the bandwidth at which the worst-case error
\begin{equation}
p_e^{\mathrm{worst}}=\frac{1-\sigma_3}{2}
\label{eq:pe_worst}
\end{equation}
reaches $3\%$. In (\ref{eq:qber_model}) the PMD-related term is the QBER of the bases actually used, i.e., the pole/equator average of Section~\ref{subsec:symmetric}, which never exceeds $p_e^{\mathrm{worst}}$; a budget defined from the worst case is therefore conservative by construction, and it provides a guarantee: within $\Delta\lambda_{3\%}$, no input state measured in its matched basis exceeds the $3\%$ projection-error level.

\section{Characterization of Channels and PMD Mitigation}
\label{sec:exper}

We apply the framework of Section~\ref{sec:framework} to deployed fiber links and demonstrate that the band-averaged rotation matrix~(\ref{eq:Mdef}) and its singular value decomposition~(\ref{eq:svd}) provide a compact, operationally meaningful characterization of broadband polarization channels. We then use the same characterization to guide a simple PMD compensation experiment. In summary, $\mathsf{R}(\lambda)$ is reconstructed from the channel measurements, and $\mathsf{M}(\Delta\lambda)$ is formed and decomposed by (\ref{eq:svd}) for a nested family of filtering bandwidths centered at $\lambda_0$. The decompositions give the polar, equatorial and worst-case error probabilities versus $\Delta\lambda$, the optimal input states and projectors, the $\sigma$-signature and the filtering budget $\Delta\lambda_{3\%}$, which together guide the choice of operating bandwidth and PMD compensation.

\subsection{Measurement Technique}
\label{subsec:measurement}

The polarization transfer of each channel is measured with the standard frequency-domain technique~\cite{williams-pmd}. The setup consists of a tunable laser (EXFO OSICS mainframe with a T100 1310~nm module), a polarization controller preparing the input states -- electrically rotated waveplates between fiber collimators -- the fiber span under test, and a polarimeter (Thorlabs PAX1000IR) connected in series. For a set of known input states the wavelength-resolved output Stokes vectors reconstruct the channel rotation $\mathsf{R}(\lambda)$ of (\ref{eq:transformation}) across the scanned interval. Measurements were performed on deployed underground fiber links in Masdar City, Abu Dhabi, around $\lambda_0 = 1310$~nm, scanning from 1300~nm to 1320~nm in $0.25$~nm steps. A full scan of one channel takes several minutes, far shorter than the hours-long timescale on which the polarization transformation of these links drifts. The polarization data (Stokes vectors) were measured for six input states, $H$, $V$, $D$, $A$, $R$ and $L$, for each channel. Only three mutually orthogonal inputs are required, so the six measured states provide substantial redundancy: they furnish eight orthonormal triads, one for each choice of $H$ or $V$, $D$ or $A$, $R$ or $L$, and the rotation stack, the band-averaged matrix and hence the entire singular-value analysis can be carried out separately from each. Repeated from different triads, the analysis reproduces the error probability to a few parts in $10^{3}$ across all 12 channels (supplementary material, Section~S2). The reconstruction algorithm and all numerical procedures are documented alongside the code (see Data and Code Availability).

From the reconstructed $\mathsf{R}(\lambda)$ the channel can also be described in conventional PMD terms. Differentiating the polarization transformation with respect to frequency yields the PMD vector $\bm{\tau}(\omega)$~\cite{gordon-kogelnik,jopson-mmm}, defined through
\begin{equation}
\frac{d\bm{s}_{\mathrm{out}}}{d\omega} = \bm{\tau}(\omega)\times\bm{s}_{\mathrm{out}},
\label{eq:pmd_vector}
\end{equation}
i.e.\ the instantaneous rotation axis about which an output state turns on the Poincar\'e sphere as the frequency varies, its magnitude $|\bm{\tau}|$ being the local DGD. A wavelength-independent $\bm{\tau}$ corresponds to the first-order PMD approximation; its variation with $\omega$ encodes the higher orders. The trajectory-based analysis retains all orders without computing $\bm{\tau}(\omega)$ at all. For a chosen filtering bandwidth $\Delta\lambda$ centered at $\lambda_0$ we form the band-averaged rotation matrix $\mathsf{M}(\Delta\lambda) = \langle\mathsf{R}(\lambda)\rangle_{\Delta\lambda}$ directly from $\mathsf{R}(\lambda)$ by numerical integration and compute its singular value decomposition~(\ref{eq:svd}). Crucially, $\mathsf{M}$ is built from $\mathsf{R}(\lambda)$ itself rather than from its frequency derivative, avoiding the noise-sensitive numerical differentiation that the PMD-vector route requires. The polar, equatorial, and worst-case error probabilities, $(1-\sigma_1)/2$, $(1-\sigma_2)/2$ and $(1-\sigma_3)/2$, follow as functions of $\Delta\lambda$ by repeating the decomposition for a family of nested bandwidths with the same central wavelength.

Throughout, we restrict attention to projection error probabilities below ${\sim}10\%$: larger error rates preclude secret-key generation in BB84-type protocols, whose asymptotic threshold is ${\approx}11\%$~\cite{shor-preskill}.

\subsection{Two Representative Channels}
\label{subsec:two_channels}

Deployed channels differ markedly in their PMD content (supplementary material, Section~S1). A pure first-order channel has the signature $(\sigma_1,\sigma_2,\sigma_3) = (1, c, c)$ -- the polar state coincides with the principal state and is undistorted ($\sigma_1=1$), while the transverse plane is contracted isotropically ($\sigma_2=\sigma_3$). Higher-order PMD lifts this degeneracy, both by reducing $\sigma_1$ below unity (the principal state itself wanders with wavelength) and by splitting $\sigma_2$ from $\sigma_3$. The departure of the measured triple from $(1,c,c)$ is therefore a direct, basis-independent measure of the higher-order contribution.

As a reference baseline we also plot the first-order PMD prediction
\begin{equation}
\bar{p}_e^{\,(1)} = \frac{(\Delta\theta)^2}{48},
\label{eq:first_order_law}
\end{equation}
derived for pure first-order PMD in our earlier work~\cite{rodimin-pra}, where $\Delta\theta = \langle\tau\rangle\,\Delta\omega$ is the accumulated rotation angle across the filtered sub-band and $\langle\tau\rangle$ its mean DGD. Equation~(\ref{eq:first_order_law}) is the equatorial $p_e$ that would result if only the mean DGD were retained from the full PMD vector; its departure from the measured SVD curve thus quantifies the higher-order content. In the figures the reference is evaluated for each sub-band from that sub-band's own mean DGD.

\begin{figure*}[!t]
\centering
\includegraphics[width=0.75\textwidth]{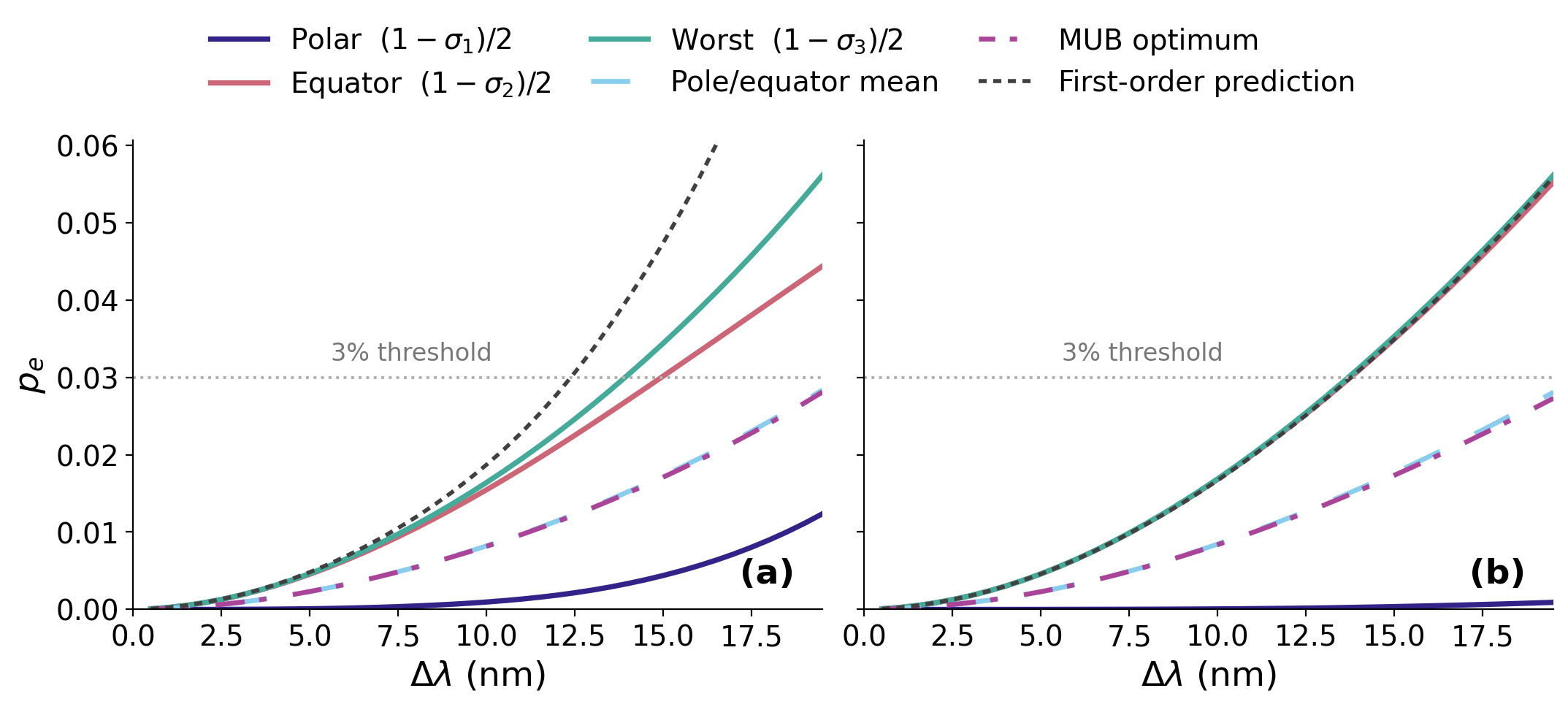}
\caption{Projection error probability $p_e$ of two deployed fiber links in Masdar City at $\lambda_0 = 1310$~nm as a function of filtering bandwidth. (a) Channel~4, showing a strong higher-order PMD contribution ($\sigma_1<1$, $\sigma_2>\sigma_3$). (b) Channel~1, showing a small higher-order PMD contribution ($\sigma_1\approx1$, $\sigma_2\approx\sigma_3$). Curves: polar $(1-\sigma_1)/2$ (dark blue, solid), equatorial $(1-\sigma_2)/2$ (red, solid), worst-case $(1-\sigma_3)/2$ (teal, solid), their pole/equator mean (light blue, solid), the optimal MUB pair given by (\ref{eq:mub_opt}) (purple, dash-dotted), and the first-order prediction, (\ref{eq:first_order_law}) (black, dotted); the horizontal dotted line marks the $3\%$ threshold. The pole/equator mean and the $45^\circ$ symmetric-pair average given by (\ref{eq:mub_opt}), nearly coincide on this scale.} 
\label{fig:combined}
\end{figure*}

Of the 12 characterized channels (supplementary material, Table~S1), Channels~1 and~4 form a matched pair: over the full $20$~nm scan their worst-case error probabilities coincide at $(1-\sigma_3)/2 = 0.058$; their filtering budgets are almost the same, $\Delta\lambda_{3\%} = 13.7$ and $13.9$~nm, respectively. Their higher-order content, however, differs markedly. Channel~1 (Fig.~\ref{fig:combined}(b)) is essentially first order: $\sigma_1 = 0.998$, the transverse plane is contracted almost isotropically ($\sigma_2-\sigma_3 = 0.002$), and the equatorial curve tracks (\ref{eq:first_order_law}) over the scanned bandwidths. Channel~4 (Fig.~\ref{fig:combined}(a)) carries a clear higher-order contribution: the principal state itself wanders with wavelength ($\sigma_1 = 0.974$), the transverse degeneracy is lifted ($\sigma_2-\sigma_3 = 0.024$), and the equatorial curve departs from (\ref{eq:first_order_law}) as the bandwidth grows.

The pair isolates what higher-order PMD does. As shown in the Appendix, rotation of the PSP axis at fixed DGD strictly lowers the equatorial error probability, whereas DGD-magnitude variation at fixed axis typically raises it slightly. An ensemble simulation with a standard random-coupling model~\cite{galtarossa} (supplementary material, Section~S3) locates the crossover between the two mechanisms at the PSP bandwidth -- the spectral interval over which the PMD vector itself remains substantially unchanged, set by the width of its autocorrelation function~\cite{karlsson-brentel}. For Channel~4, the PSP bandwidth is about 7~nm: for $\Delta\lambda < 7$~nm the PMD is first-order, with the error probability following the quadratic curve. Beyond the 7~nm bandwidth, the PSP axis wander of Channel~4 lowers the equatorial value below the first-order prediction. For Channel~1 the PSP bandwidth spans almost the entire scanning interval, so higher-order PMD effects are barely observable.

\subsection{PMD Compensation by Channel Concatenation}
\label{subsec:compensation}

\begin{figure*}[!t]
\centering
\includegraphics[width=0.62\textwidth]{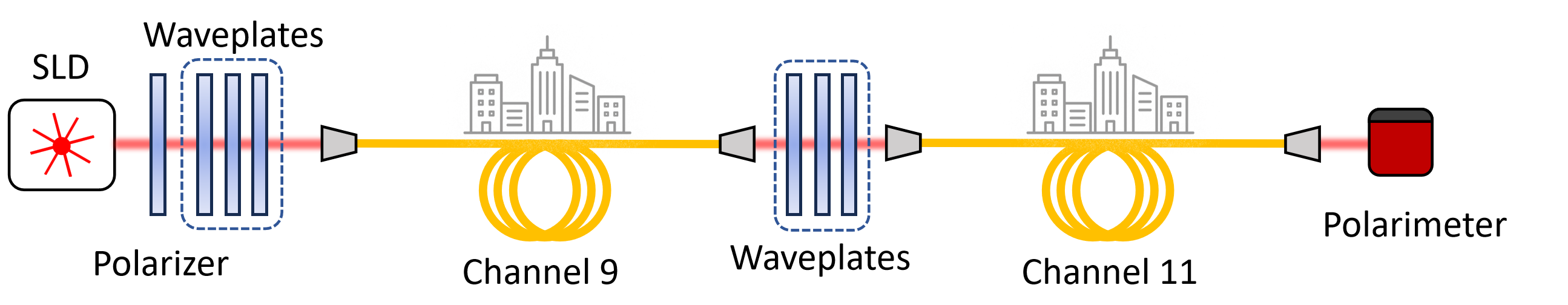}
\caption{Experimental layout of the PMD compensation. The first set of waveplates prepares the input polarization state, while the second sets the relative orientation of the PMD vectors of the two channels. SLD: superluminescent diode.}
\label{fig:setup_comp}
\end{figure*}

A frequency-independent polarization rotation inserted between two fiber spans sets the mutual orientation of their PMD vectors in the concatenation rule~\cite{gordon-kogelnik}; anti-aligning them makes the two DGDs subtract, so that spans of comparable DGD cancel much of one another's first-order PMD. For the demonstration, we selected two strongly depolarizing channels, Channels~9 and~11, whose measured PMD-vector magnitudes are comparable. The two channels were joined through a polarization controller that sets the relative orientation of the two PMD vectors.

The experimental layout is shown in Fig.~\ref{fig:setup_comp}. With a broadband superluminescent-diode (SLD) source, filtered to $12$~nm, at the input of the concatenated link, the polarization controller between the two channels was adjusted to bring the combined output toward the lowest-$p_e$ configuration. That is, the degree of polarization of the band-filtered output was maximized --- equivalently, $\sigma_1$ of the combined channel was driven toward unity. The resulting channel was then characterized with the same wavelength-resolved technique.

\begin{figure}[!t]
\centering
\includegraphics[width=0.9\columnwidth]{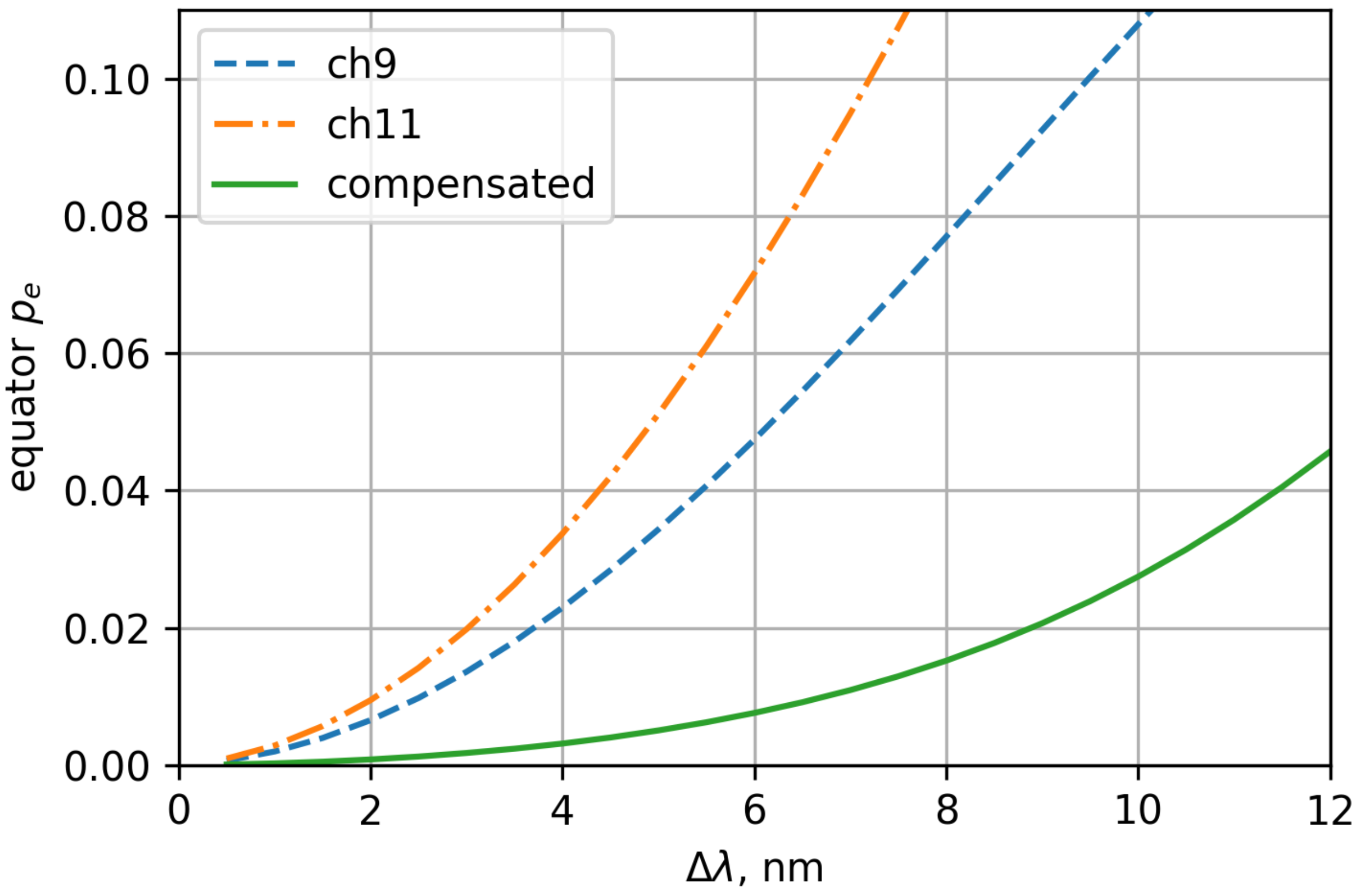}
\caption{PMD compensation by concatenation of two strongly depolarizing channels, joined through a polarization controller set with the broadband SLD source: equatorial $p_e$ of Channel~9 (blue, dashed), Channel~11 (orange, dash-dotted) and the compensated cascade (green, solid). The $p_e$ of the compensated cascade is reduced relative to the constituent channels by about a factor of six, despite twice the fiber length.}
\label{fig:compensation}
\end{figure}

Fig.~\ref{fig:compensation} shows $p_e$ of the compensated, concatenated channel together with that of the constituent channels. The concatenation reduces the broadband error probability relative to the individual channels: at 5~nm filtering bandwidth the separate channels give equatorial $p_e$ of $3.2\%$ and $4.8\%$, whereas the concatenated channel -- twice the length -- gives an equatorial $p_e$ of $0.5\%$, at least a sixfold reduction relative to the better of the two constituents.

\section{Discussion and Summary}
\label{sec:discussion}

We have presented a method for characterizing a fiber-based polarization channel carrying broadband quantum light. The geometric optimization over the Poincar\'e sphere reduces to a single decomposition of the band-averaged rotation matrix $\mathsf{M}(\Delta\lambda)=\langle\mathsf{R}(\lambda)\rangle$. The optimal input states, the mutually unbiased measurement bases, and their error probabilities all follow in closed form from its SVD.

The presented method avoids the noise associated with numerical differentiation in conventional PMD analysis by obtaining $\mathsf{M}$ directly as the wavelength average of $\mathsf{R}(\lambda)$. The error probability reconstructed from eight different input triads of the same measurement reproduces to a few parts in $10^3$ across all 12 channels (supplementary material, Section~S2).

As a channel characterization, the method provides two descriptors: the filtering budget $\Delta\lambda_{3\%}$ and the structural higher-order fingerprint -- the singular-value signature $(\sigma_1,\sigma_2,\sigma_3)$. Together, they summarize both how much a channel may be filtered and what kind of PMD it carries. The two descriptors are stable over time, because they depend only on the wavelength dependence of the channel, which remains stable over month-long periods in buried fibers~\cite{brodsky-hinge}. The fast, wavelength-independent polarization drift typically arises from short exposed fiber sections and can be compensated by standard polarization tracking~\cite{oe2025}.

These descriptors are directly useful for the choice of PMD mitigation strategy. Higher-order PMD can play a favorable role in passive operation, since it statistically reduces the error probability below the first-order prediction. However, for active PMD mitigation with a birefringent element, it is preferable to work with a channel dominated by first-order PMD, since anti-alignment then cancels the two vectors most completely.

The protocol architecture determines the details of the compensation. In a prepare-and-measure link, compensation is always local: a polarization controller and a matched birefringent element (e.g., a polarization-maintaining fiber patch tuned mechanically or thermally) supply the opposing PMD vector in series, as demonstrated by the concatenation experiment of Section~\ref{subsec:compensation}. In entanglement-based QKD with two separate channels, compensation can also be done nonlocally, by aligning one arm's PMD vector with the other's via a polarization controller in that arm~\cite{antonelli-prl,shtaif-oe}. Our concatenation experiment effectively demonstrates the feasibility of compensation with two deployed channels: the only difference is that in the nonlocal case the two-photon state sees one arm reversed relative to the cascaded configuration.

The framework also suggests an alternative measurement modality that dispenses with the tunable laser altogether. Because the polarimeter reading of band-filtered broadband light is itself the band-averaged Stokes vector, the central object of the analysis is directly measurable: launching three known, orthogonal input states from a broadband SLD source and recording the full output Stokes vectors behind a bandpass filter determines $\mathsf{M}$ at that filter width, with no wavelength scan and no reconstruction of $\mathsf{R}(\lambda)$. Three such measurements fix all nine entries of $\mathsf{M}$ exactly. A set of switchable filters with increasing bandwidth then samples the entire $p_e$-versus-bandwidth analysis at the discrete filter widths, replacing the swept-wavelength measurement. The compensation experiment of Section~\ref{subsec:compensation}, in which the polarization controller was set by maximizing the degree of polarization of SLD light behind a $12$~nm filter, is an instance of this modality.

Finally, the $p_e$-versus-bandwidth curve supplies the ingredient needed to optimize a quantum-key-distribution link for secret key rate: filtering trades photon flux against quantum bit error rate, and the measured pole/equator average $p_e(\Delta\lambda)$ makes this trade-off quantitative for a given deployed channel, so it can be folded into a key-rate model to choose the operating bandwidth. A full key-rate optimization is left to future work.

\appendix[Two Higher-Order Mechanisms: Covariance Form and PSP Wander]
\label{app:wander}

The two higher-order mechanisms -- variation of the DGD magnitude at a fixed PSP axis, and rotation (``wander'') of the PSP axis with wavelength -- admit a single unifying expression, obtained by carrying the band average to second order in the accumulated rotation angle.

\subsection{Second-Order Expansion}
\label{app:expansion}
Writing $[\bm{a}]_\times$ for the skew-symmetric cross-product matrix of a vector $\bm{a}$, $[\bm{a}]_\times\bm{x}=\bm{a}\times\bm{x}$, the PMD-vector definition~(\ref{eq:pmd_vector}) becomes a matrix differential equation for the channel rotation~\cite{gordon-kogelnik},
\begin{equation}
\frac{d\mathsf{R}}{d\omega}=[\bm{\tau}(\omega)]_\times\,\mathsf{R}(\omega),\qquad \mathsf{R}(\omega_0)=\mathsf{I},
\label{eq:R_ode}
\end{equation}
where the initial condition removes the constant rotation at $\omega_0$, which does not affect the singular values. Its solution can be written as a single exponential, the Magnus expansion~\cite{magnus},
\begin{equation}
\mathsf{R}(\omega)=\exp\bigl([\bm{\theta}+\bm{\eta}+\cdots]_\times\bigr),
\label{eq:magnus}
\end{equation}
\begin{align}
\bm{\theta}&=\int_{\omega_0}^{\omega}\bm{\tau}(\omega')\,d\omega',\nonumber\\
\bm{\eta}&=\frac{1}{2}\int_{\omega_0}^{\omega}\!d\omega'\int_{\omega_0}^{\omega'}\!d\omega''\;\bm{\tau}(\omega')\times\bm{\tau}(\omega''),
\label{eq:magnus_terms}
\end{align}
where $\bm{\theta}$ is the accumulated rotation vector (first order in $\tau\,\Delta\omega$) and $\bm{\eta}$ the non-commutativity correction (second order); the vector form follows from $\bigl[[\bm{a}]_\times,[\bm{b}]_\times\bigr]=[\bm{a}\times\bm{b}]_\times$. The omitted terms are nested cross products of third and higher order. For a wavelength-independent PSP axis all terms beyond $\bm{\theta}$ vanish identically and (\ref{eq:magnus}) reduces to the first-order rotation of Section~\ref{sec:framework}. The Magnus expansion has previously been applied to polarization transfer matrices, including the case of polarization-dependent loss~\cite{reimer-mueller}, and the accuracy of its truncation as a function of optical bandwidth has been established~\cite{reimer-magnus}.

Expanding (\ref{eq:magnus}) to second order in $\theta$,
\begin{equation}
\mathsf{R}(\omega)=\mathsf{I}+[\bm{\theta}]_\times+[\bm{\eta}]_\times+\tfrac12[\bm{\theta}]_\times^{2}+O(\theta^{3}),
\label{eq:Rexp}
\end{equation}
where the mixed term $\tfrac12\{[\bm{\theta}]_\times,[\bm{\eta}]_\times\}$ and $\tfrac12[\bm{\eta}]_\times^{2}$ are of third and fourth order and are absorbed in the remainder. Averaging over the band, and using that the linearity of $[\cdot]_\times$ lets the average pass inside the bracket for the linear terms but not for the quadratic one, gives
\begin{equation}
\mathsf{M}=\mathsf{I}+[\langle\bm{\theta}\rangle]_\times+[\langle\bm{\eta}\rangle]_\times+\tfrac12\big\langle[\bm{\theta}]_\times^{2}\big\rangle+O(\theta^{3}).
\label{eq:magnus_avg}
\end{equation}
The aim is now to separate the part of $\mathsf{M}$ that is a pure rotation from the part that contracts the Poincar\'e sphere, and to analyze only the latter: a rotation costs no fidelity, whereas the contraction sets the error probabilities. In (\ref{eq:magnus_avg}) the two linear terms are skew-symmetric and generate a rotation. The quadratic term is symmetric, but it is not by itself the contraction: a rotation expanded to second order carries its own symmetric part, $\tfrac12[\langle\bm{\theta}\rangle]_\times^{2}$, which must be removed first. What remains, the excess of $\langle[\bm{\theta}]_\times^{2}\rangle$ over $[\langle\bm{\theta}\rangle]_\times^{2}$, alone depolarizes, and it is a covariance: with the identity $[\bm{a}]_\times^{2}=\bm{a}\bm{a}^{\mathsf T}-|\bm{a}|^{2}\mathsf{I}$, the excess of a mean square over the square of the mean gives, from the outer product, the covariance tensor $\mathsf{C}$ of $\bm{\theta}$ across the band and, from the norm, its trace:
\begin{align}
\mathsf{C}&\equiv\langle\bm{\theta}\bm{\theta}^{\mathsf T}\rangle-\langle\bm{\theta}\rangle\langle\bm{\theta}\rangle^{\mathsf T}=\mathrm{Cov}_\omega[\bm{\theta}],\nonumber\\
\big\langle[\bm{\theta}]_\times^{2}\big\rangle&-[\langle\bm{\theta}\rangle]_\times^{2}=\mathsf{C}-(\mathrm{tr}\,\mathsf{C})\,\mathsf{I}.
\label{eq:cov_tensor}
\end{align}
Since $\mathsf{I}+[\langle\bm{\theta}\rangle+\langle\bm{\eta}\rangle]_\times+\tfrac12[\langle\bm{\theta}\rangle]_\times^{2}$ agrees with the rotation $\mathsf{R}(\langle\bm{\theta}\rangle+\langle\bm{\eta}\rangle)$ to $O(\theta^{3})$, (\ref{eq:magnus_avg}) becomes
\begin{equation}
\mathsf{M}=\mathsf{R}(\langle\bm{\theta}\rangle+\langle\bm{\eta}\rangle)\Bigl[\mathsf{I}-\tfrac12\bigl((\mathrm{tr}\,\mathsf{C})\,\mathsf{I}-\mathsf{C}\bigr)\Bigr]+O(\theta^{3}),
\label{eq:M_polar}
\end{equation}
a pure rotation times a symmetric contraction: the polar decomposition of $\mathsf{M}$ to second order. The singular values of $\mathsf{M}$ are the eigenvalues of the contraction, $\sigma_i=1-\tfrac12(c_j+c_k)$, with $c_1\ge c_2\ge c_3$ the eigenvalues of $\mathsf{C}$ and $\{i,j,k\}$ a permutation of $\{1,2,3\}$, and the input states $\bm{v}_i$ are the eigenvectors of $\mathsf{C}$. Hence
\begin{equation}
p_i=\frac{1-\sigma_i}{2}=\frac{c_j+c_k}{4},
\label{eq:cov_form}
\end{equation}
where $p_1$, $p_2$ and $p_3$ are the polar, equatorial and worst-case error probabilities of Section~\ref{sec:framework}. The non-commutativity correction $\bm{\eta}$ enters only the rotation factor; its fluctuations would contribute to $\mathsf{C}$ at third order.

Summing (\ref{eq:cov_form}) over the three directions,
\begin{equation}
\sum_i p_i=\frac{\mathrm{tr}\,\mathsf{C}}{2}=\frac{\langle|\bm{\theta}|^{2}\rangle-|\langle\bm{\theta}\rangle|^{2}}{2},
\label{eq:p_sum}
\end{equation}
so the total error probability over the three input directions $\bm{v}_i$ is set by the total spread of the rotation vector across the band and is independent of how that spread is oriented; the anisotropy of $\mathsf{C}$ only decides how the total is shared, with the direction of least variance receiving the largest share.

Equation~(\ref{eq:cov_form}) contains both higher-order mechanisms as limiting geometries of the curve $\bm{\theta}(\omega)$ and hence of $\mathsf{C}$.

\subsection{DGD-Magnitude Variation at a Fixed Axis}
\label{app:magnitude}
If the PSP axis $\hat{\bm{n}}$ is fixed, the rotation vector keeps a fixed direction, $\bm{\theta}(\omega)=\theta(\omega)\,\hat{\bm{n}}$ with the signed accumulated angle $\theta(\omega)=\int_{\omega_0}^{\omega}\tau\,d\omega'$ (the output states then trace circles about $\hat{\bm{n}}$ on the Poincar\'e sphere, a great circle for the equatorial input), so $\mathsf{C}$ has rank one: $c_1=\mathrm{Var}_\omega[\theta]$ and $c_2=c_3=0$. Equation~(\ref{eq:cov_form}) then gives $\sigma_1=1$ --- the axis itself is undistorted --- and a degenerate transverse pair,
\begin{equation}
p_2=p_3=\frac{\mathrm{Var}_\omega[\theta]}{4}.
\label{eq:varform}
\end{equation}
The error probability is thus set by the spread of the accumulated angle across the band, not by the band-averaged DGD alone. A constant DGD gives $\mathrm{Var}_\omega[\theta]=(\Delta\theta)^2/12$ and recovers the first-order law~(\ref{eq:first_order_law}). Redistributing the same $\langle\tau\rangle$ across the band moves $p_2$ in either direction, since the state lingers where the DGD is small: DGD concentrated near the band center parks the state near the two ends of its arc and raises $p_2$, DGD pushed to the band edges keeps it near the middle and lowers $p_2$.

\subsection{PSP Wander at Constant DGD}
\label{app:psp_wander}
If instead the DGD is constant, $\bm{\theta}(\omega)$ is a curve traversed at constant speed, $|d\bm{\theta}/d\omega|=\tau_0$. Writing the total variance as a mean over pairs of band points, $\mathrm{tr}\,\mathsf{C}=\tfrac12\big\langle\!\big\langle|\bm{\theta}(\omega)-\bm{\theta}(\omega')|^2\big\rangle\!\big\rangle$, and using that a chord never exceeds its arc, $|\bm{\theta}(\omega)-\bm{\theta}(\omega')|\le\tau_0|\omega-\omega'|$,
\begin{equation}
\mathrm{tr}\,\mathsf{C}\le\frac{\tau_0^2\Delta\omega^2}{12}=\frac{(\Delta\theta)^2}{12},
\label{eq:chord_arc}
\end{equation}
with equality if and only if the axis is fixed. Two corollaries follow. The sum rule~(\ref{eq:p_sum}) is bounded, $\sum_i p_i\le(\Delta\theta)^2/24$: at constant DGD, wander lowers the total. And since $c_2\ge0$,
\begin{equation}
p_2=\frac{c_1+c_3}{4}\le\frac{\mathrm{tr}\,\mathsf{C}}{4}\le\frac{(\Delta\theta)^2}{48},
\label{eq:downward}
\end{equation}
so wander can only move the equatorial error probability below the first-order law. Part of the reduced total is moved into the polar direction, $\sigma_1<1$; wander with a preferred plane additionally splits $c_2$ from $c_3$ and hence $\sigma_2$ from $\sigma_3$.

\subsection{Reading the Signature}
\label{app:diagnostic}
The signature $(\sigma_1,\sigma_2,\sigma_3)$ therefore reads as a diagnostic of the channel's PMD character: $\sigma_1<1$ arises only through axis wander; $\sigma_1=1$ with a degenerate pair $\sigma_2=\sigma_3$ off the first-order law, in either direction, indicates pure DGD-magnitude structure; $\sigma_2\neq\sigma_3$ indicates anisotropic wander. The first of these is exact, not perturbative: for a fixed rotation axis every $\mathsf{R}(\lambda)$, and hence $\mathsf{M}$, leaves the axis vector invariant, so $\sigma_1=1$ at any band angle. The quantitative relations~(\ref{eq:cov_form})--(\ref{eq:downward}) hold in the small-angle regime; the wander classification of the strongly depolarizing channels of Table~S1, whose full-band angles exceed it, rests on the exact statement alone.

Which mechanism dominates in a given deployed link is a statistical question, addressed in Section~S3 of the supplementary material by an ensemble simulation with a standard random-coupling model. Below the PSP bandwidth the axis is effectively frozen and the magnitude mechanism biases the equatorial error probability slightly above the first-order law; beyond it the wander mechanism dominates and the deviation is downward --- the typical broadband regime, and the one observed for Channel~4 (Fig.~\ref{fig:combined}(a)). The channel-by-channel classification of the 12 measured links is given in Section~S1 of the supplementary material.

\section*{Acknowledgment}
During the preparation of this work the authors used Anthropic's Claude to assist in formalizing the authors' heuristic results into the mathematical framework presented here and in editing the authors' original text. The tool was not used to generate research data; all derivations were independently verified by the authors, who take full responsibility for all content. The authors declare no conflicts of interest.

\section*{Data and Code Availability}
The wavelength-resolved polarization measurements of all 12 deployed channels and of the compensation experiment are openly available at \url{https://github.com/a-ponasenko/broadband-pmd}. The Python library implementing the reconstruction of $\mathsf{R}(\lambda)$, the band-averaged matrix analysis and the Poincar\'e-sphere visualization, together with a notebook reproducing the figures of this paper and of the supplementary material, is openly available in the same repository under the MIT license. 

\bibliographystyle{IEEEtran_jlt} 
\bibliography{refs_jlt}

\end{document}